\documentclass[a4paper,12pt]{article}
\usepackage[ansinew]{inputenc}
\usepackage[T1]{fontenc}
\usepackage[english]{babel}
\usepackage{latexsym}
\usepackage{amsmath}
\usepackage{amssymb}
\usepackage[dvips]{graphicx}
\usepackage{epsfig}
\usepackage{subfigure}
\newcommand{\mc}{\mathcal}

\newcommand{\eps}{\varepsilon}

\newcommand{\beq}{\begin{equation}}
\newcommand{\eeq}{\end{equation}}
\newcommand{\beqa}{\begin{eqnarray}}
\newcommand{\eeqa}{\end{eqnarray}}
\newcommand{\beqan}{\begin{eqnarray*}}
\newcommand{\eeqan}{\end{eqnarray*}}
\newcommand{\nn}{\nonumber}

\newcommand{\bc}{\begin{center}}
\newcommand{\ec}{\end{center}}

\title{Baryon asymmetry and dark matter from soft leptogenesis}
\author{Heidi Kuismanen {\small and} Iiro Vilja  \\
 {\small  \textit{Department of Physics and Astronomy, 
University of Turku, 20014 Turku, Finland}}}
\date{\today}

\begin{document}

\maketitle

\begin{abstract}
The framework for soft leptogenesis minimally extended with a DM sector is studied. A heavy singlet neutrino superfield acts as the source for (s)lepton asymmetry and by coupling to the singlet DM superfield it produces a DM particle density through decays. The nature of DM generated is twofold depending on whether the Yukawa and DM couplings are either small or large. With sufficiently small Yukawa and DM couplings DM annihilations into MSSM particles are slow and as a consequence all DM particles form the DM component. The solutions to Boltzmann equations are given and the dependence between the DM masses and coupling are presented in this weak coupling regime. Also, the behavior of the efficiency of producing asymmetric DM is determined with weak couplings. We note that a different outcome arises if the couplings are larger because then the ADM component is dominant due to the effectiveness of DM decays into the MSSM sector. 
\end{abstract}

\section{Introduction}

Our current understanding of the universe is that around 5 $\%$ of the energy density belongs to ordinary baryonic matter while the rest stems from dark matter (DM) and dark energy with proportions 22 $\%$ and 73 $\%$, respectively \cite{wmap}. The existence of all three components is more or less a mystery and points towards physics beyond the Standard Model (SM). The fact that there seems to be almost exclusively matter and no antimatter in the universe suggests there is matter-antimatter asymmetry. Since the SM cannot accommodate sufficient baryogenesis on its own, baryogenesis mechanisms have been constructed in various Grand Unified Theory (GUT) scenarios. Alongside GUT models, baryogenesis via leptogenesis \cite{yanagida} has been studied extensively over the last twenty years or so. 

As for DM, a viable candidate arises in the Minimal Supersymmetric Standard Model (MSSM), namely the Lightest Supersymmetric Particle (LSP) \cite{LSPold,lspreferences}. Usually, the lightest neutralino (i.e.\ a combination of the superpartners of the neutral gauge bosons and higgs sparticles), or in supergravity theories the gravitino, assumes the role of the LSP, and the success of the LSP in producing the correct DM abundance depends on the adjustment of parameters within the MSSM. However, since the DM and baryonic energy densities are close to each other, $\Omega_\text{DM}/\Omega_\text{b}\sim 5$, it is natural to assume that DM and baryogenesis are somehow related. The connection between baryogenesis and DM has also been widely explored \cite{admreferences}. Moreover, the possibility of the DM energy density being the product of an asymmetry has led to the concept of asymmetric DM (ADM) where DM and DM antiparticles are distinct from another and produce an asymmetry analogous to baryons. Note, however, that as ADM scenarios incorporate MSSM, also LSP is always included in the ADM models but is not necessarily good candidate for DM.

Over the years various realizations of leptogenesis have been devised \cite{fps}-\cite{softlofti}. They are usually motivated by SO(10) scenarios where $B-L$ is a gauge symmetry, and it is natural to have a heavy right-handed SM singlet neutrino that can also be responsible for light neutrino masses via the seesaw mechanism \cite{seesaw,seesawtoka}. Adding a heavy singlet neutrino superfield into the Minimal Supersymmetric SM (MSSM) gives rise to soft leptogenesis where the soft breaking of supersymmetry allows the scalar superpartner of the heavy neutrino to generate the (s)lepton asymmetry \cite{grosskasnirrou,ambgiurai}. In contrast to non-SUSY models, only one family of heavy singlets is required.   

Furthermore, various scenarios incorporating leptogenesis and DM have been introduced \cite{cosme}-\cite{chun}. In this article, we investigate a model that encompasses soft leptogenesis and DM generation from the same heavy singlet superfield. Previously a model with $SU(2)_L$ triplet DM and supersymmetric leptogenesis has been studied \cite{eungjinchun}. Our model is a minimal extension to the Minimal Supersymmetric Standard Model (MSSM) and it consists of MSSM augmented with a single SM singlet DM superfield and a heavy singlet neutrino superfield that enables soft leptogenesis after soft SUSY breaking. In section \ref{malli} we introduce our model and present the CP violation parameters, in section \ref{boltzmann} we solve the Boltzmann equations and the resulting baryon and DM abundances, and in section \ref{paatelmat} we present our conclusions.   

%needed if the LSP paradigm within MSSM does not work

\section{The model}
\label{malli}

Models incorporating leptogenesis and DM generation have the potential to produce the observed matter and DM energy densities and offer a possible explanation as to why the abundances are similar in magnitude. So far the MSSM has been considered as one of the most appealing extension to the SM, and with the addition of a heavy singlet (s)neutrino, soft leptogenesis can arise. To account for DM, the soft leptogenesis framework can be extended with an additional DM sector. We have augmented the MSSM with a singlet superfield consisting of a heavy neutrino and sneutrino and a singlet DM superfield. The heavy (s)neutrino couples to the MSSM (s)leptons, Higgses and higgsinos, which gives rise to the soft leptogenesis scenario. The DM sector couples to the heavy (s)neutrino as well and this provides additional decay channels for the singlet (s)neutrino and a source for DM production through these decays. The superpotential is
\beqa
W=W_\text{MSSM}+y_\alpha NL_\alpha H_u+\frac{1}{2}MNN+m_\Phi \Phi\Phi ^c+\frac{1}{2}\lambda N\Phi\Phi.
\eeqa
The discrete symmetry $Z_2$ imposed on $\Phi$ forbids terms $\sim \Phi^3$ and $\sim NN\Phi$ . After adding soft supersymmetry breaking terms we get the part of the Lagrangian that includes the soft leptogenesis and darkgenesis interactions
\beqa
\mc L&=&\lambda \phi N\chi+y_\alpha H_uL_\alpha N+y_\alpha\widetilde{L}_\alpha\widetilde{H}_uN+\text{h.c.}\nn\\
&&+\widetilde{N}_+\bigg[\frac{1}{\sqrt{2}}\lambda e^{-i\theta_b/2}\chi\chi+\frac{1}{\sqrt 2}y_\alpha e^{-i\theta_b/2}L_\alpha\widetilde{H}_u+\frac{1}{2\sqrt 2}\lambda M^*e^{i\theta_b /2}\phi^2\nn\\
&&+\frac{1}{\sqrt 2}y_\alpha M^*e^{i\theta_b/2}\widetilde{L}_\alpha H_u+\frac{1}{2}c_1e^{-i\theta_b/2}\phi^2+\frac{1}{2}c_2e^{-i\theta_b/2}\widetilde{L}_\alpha H_u+\text{h.c.}\bigg]\nn\\
&&+\widetilde{N}_-\bigg[\frac{i}{\sqrt 2}\lambda e^{-i\theta_b/2}\chi\chi+\frac{i}{\sqrt 2}y_\alpha e^{-i\theta_b/2}L_\alpha \widetilde{H}_u-\frac{i}{2\sqrt 2}\lambda M^* e^{i\theta_b/2}\phi^2\nn\\
&&-\frac{i}{\sqrt 2}y_\alpha M^*e^{i\theta_b/2}\widetilde{L}_\alpha H_u+\frac{i}{\sqrt 2}c_1e^{-i\theta_b/2}\phi^2+\frac{i}{\sqrt 2}c_2 e^{-i\theta_b/2}\widetilde{L}_\alpha H_u+\text{h.c.}\bigg],
\eeqa
where the sneutrino mass eigenstates are
\beqa
\widetilde{N}_+&=&\frac{1}{\sqrt 2}\left ( e^{i\theta _b/2}\widetilde{N}+e^{-i\theta _b/2}\widetilde{N}^*\right ),\\
\widetilde{N}_-&=&\frac{1}{\sqrt{2}i}\left ( e^{i\theta _b/2}\widetilde{N}-e^{-i\theta _b/2}\widetilde{N}^*\right ).
\eeqa
The DM particles $\chi$ and $\phi$ represent the DM fermion and scalar, respectively.

\begin{figure}[htb]
\centering
\subfigure[]{
\includegraphics[scale=0.8]{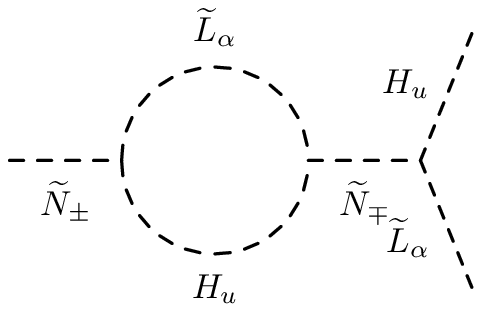}
\label{fig:subfig1}
}
\quad
\subfigure[]{
\includegraphics[scale=0.8]{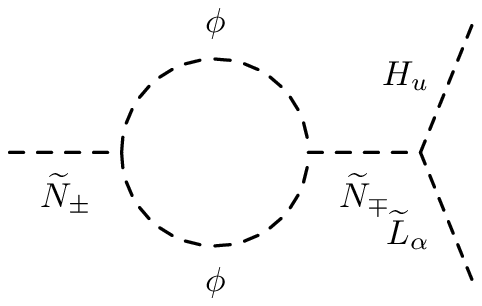}
\label{fig:subfig2}
}

\subfigure[]{
\includegraphics[scale=0.8]{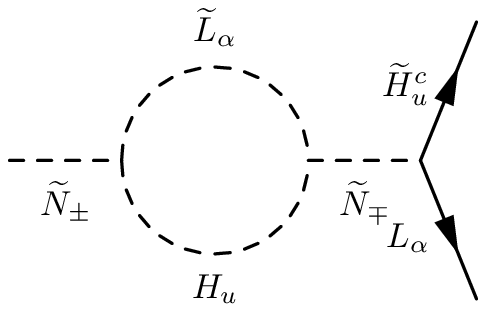}
\label{fig:subfig3}
}
\quad
\subfigure[]{
\includegraphics[scale=0.8]{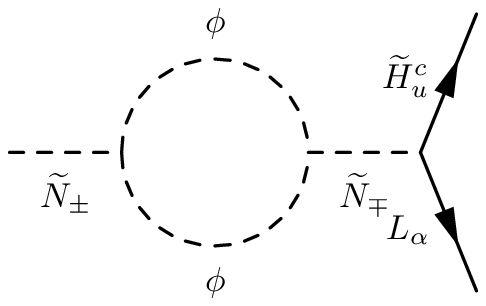}
\label{fig:subfig4}
}

\subfigure[]{
\includegraphics[scale=0.8]{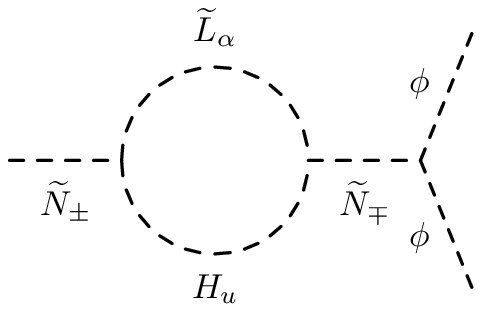}
\label{fig:subfig5}
}
\quad
\subfigure[]{
\includegraphics[scale=0.8]{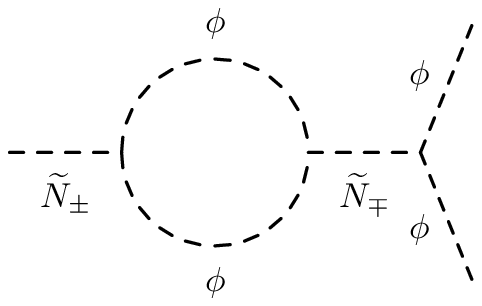}
\label{fig:subfig6}
}
\caption{The relevant loop diagrams contributing to (s)lepton and DM asymmetries.}
\label{fig:loops}
\end{figure}

The lepton and DM asymmetries arise from the self-energy loop diagrams shown in Figs. \ref{fig:subfig1}-\ref{fig:subfig6}. While the process $\widetilde{N}_\pm\rightarrow \chi\chi^c$ does not produce a DM asymmetry, i.e. an asymmetry between $\chi$ and $\chi^c$, the decays $\widetilde{N}_\pm\rightarrow \phi\phi$ and $\widetilde{N}_\pm\rightarrow \phi^*\phi^*$ can create an asymmetry between $\phi$ and $\phi^*$. This is the source for possible ADM in our model and the asymmetric component consists of scalar DM. The CP violation parameter for (s)lepton production is defined as
\beqa
\eps_L=\frac{\sum_f\left[\Gamma(\widetilde{N}_+\rightarrow f)-\Gamma(\widetilde{N}_+\rightarrow \bar f)+\Gamma(\widetilde{N}_-\rightarrow f)-\Gamma(\widetilde{N}_-\rightarrow \bar f)\right]}{\sum_f\left[\Gamma(\widetilde{N}_+\rightarrow f)+\Gamma(\widetilde{N}_+\rightarrow \bar f)+\Gamma(\widetilde{N}_-\rightarrow f)+\Gamma(\widetilde{N}_-\rightarrow \bar f)\right]},
\eeqa
where $f$ stands for the fermionic ($l_\alpha \widetilde{H}_u^c$) and bosonic ($\tilde l_\alpha H_u$) final states. The DM asymmetry is defined in a similar manner:
\beqa
\eps_\text{DM}=\frac{\Gamma\left (\widetilde{N}_+\rightarrow \phi\phi\right )-\Gamma\left (\widetilde{N}_+\rightarrow \phi^*\phi^*\right )+\Gamma\left (\widetilde{N}_-\rightarrow \phi\phi\right )-\Gamma\left (\widetilde{N}_-\rightarrow \phi^*\phi^*\right )}{\Gamma\left (\widetilde{N}_+\rightarrow \phi\phi\right )+\Gamma\left (\widetilde{N}_+\rightarrow \phi^*\phi^*\right )+\Gamma\left (\widetilde{N}_-\rightarrow \phi\phi\right )+\Gamma\left (\widetilde{N}_-\rightarrow \phi^*\phi^*\right )}
\eeqa
\newpage
The temperature-dependent CP violation parameter for the (s)leptons is
\beqa \label{epsl}
\eps_L(T)&=& \Bigg[-\frac{4}{16\pi}\frac{M_{\widetilde{N}_+}^2-M_{\widetilde{N}_-}^2}{(M_{\widetilde{N}_+}^2-M_{\widetilde{N}_-}^2)^2+\Pi_{--}^2}\frac{|M|}{v_u^2}m_\text{eff}\nn\\
&&\times\left(\frac{|M|^2}{2}-\frac{|A|^2}{2}\right)\bigg (2|A||M|\cos\left(\theta_A+\frac{\pi}{2}+\theta_M-\theta_b\right)\frac{|M|}{v_u^2}m_\text{eff}\nn\\
&&+|\lambda ||M| |c_1|\cos\left(-\theta_\lambda+ \frac{\pi}{2}+\theta_M-\theta_b+\theta_{c_1}\right)\bigg )\frac{c_B}{16\pi M_{\widetilde{N}_+}}\nn\\
&&-\frac{4}{16\pi}\frac{M_{\widetilde{N}_-}^2 - M_{\widetilde{N}_+}^2}{(M_{\widetilde{N}_-}^2 - M_{\widetilde{N}_+}^2)^2+\Pi_{++}^2}\left(-\frac{|M|^2}{2}+\frac{|A|^2}{2}\right)\frac{|M|}{v_u^2}m_\text{eff}\nn\\
&&\times\bigg (2|M||A|\cos\left(\theta_M+\theta_A-\theta_b+\frac{\pi}{2}\right)\frac{|M|}{v_u^2}m_\text{eff}\nn\\
&&+|c_1||\lambda ||M|\cos\left(\theta_{c_1}-\theta_\lambda+\theta_M-\theta_b+\frac{\pi}{2}\right)\bigg )\frac{c_B}{16\pi M_{\widetilde{N}_-}}\nn\\
&&-\frac{4}{16\pi}\frac{M_{\widetilde{N}_-}^2-M_{\widetilde{N}_+}^2}{(M_{\widetilde{N}_-}^2-M_{\widetilde{N}_+}^2)^2+\Pi_{++}^2}\frac{|M|}{v_u^2}\frac{m_\text{eff}}{2}\bigg (2|M||A|\sin\left(-\theta_M-\theta_A+\theta_b\right)\frac{|M|}{v_u^2}m_\text{eff}\nn\\
&&+|c_1||\lambda ||M|\sin\left(\theta_\lambda-\theta_{c_1}-\theta_M+\theta_b\right)\bigg )\frac{c_F}{16\pi}M_{\widetilde{N}_-}\nn\\
&&-\frac{4}{16\pi}\frac{M_{\widetilde{N}_+}^2-M_{\widetilde{N}_-}^2}{(M_{\widetilde{N}_+}^2-M_{\widetilde{N}_-}^2)^2+\Pi_{--}^2}\frac{|M|}{v_u^2}\frac{m_\text{eff}}{2}\bigg (2|A||M|\sin\left(\theta_b-\theta_A-\theta_M\right)\frac{|M|}{v_u^2}m_\text{eff}\nn\\
&&+|c_1||M||\lambda|\sin\left(\theta_b-\theta_{c_1}+\theta_\lambda-\theta_M\right)\bigg )M_{\widetilde{N}_+}\frac{c_F}{16\pi}\Bigg ]\nn\\
&&\times \Bigg[\left (|M|^2+|A|^2+2|M||A|\cos\left(\theta_M-\theta_b+\theta_A\right)\right )\frac{|M|}{v_u^2}m_\text{eff}\frac{c_B}{8\pi M_{\widetilde{N}_+}}\nn\\
&&+\left (|M|^2 + |A|^2-2|M||A|\cos\left (\theta_M-\theta_b+\theta_A\right )\right )\frac{|M|}{v_u^2}m_\text{eff}\frac{c_B}{8\pi M_{\widetilde{N}_-}}\nn\\
&&+M_{\widetilde{N}_-}\frac{|M|}{v_u^2}m_\text{eff}\frac{c_F}{8\pi}+M_{\widetilde{N}_+}\frac{|M|}{v_u^2}m_\text{eff}\frac{c_F}{8\pi}\Bigg ]^{-1},
\eeqa
where we see the influence of the DM sector couplings in the $\widetilde{N}_\pm$ self-energies and $\widetilde{N}_\pm\rightarrow \widetilde{N}_\mp$ transitions. We have used $c_2\equiv A\sum_\alpha y_\alpha$ and $m_\text{eff}=v_u^2\sum_\alpha |y_\alpha|^2/M$. The net $\phi$ asymmetry is
\beqa
\eps_\text{DM}(T)&=&\Bigg[-\frac{4}{16\pi}\frac{M_{\widetilde{N}_+}^2-M_{\widetilde{N}_-}^2}{(M_{\widetilde{N}_+}^2-M_{\widetilde{N}_-}^2)^2+\Pi_{--}^2}\left(\frac{|M|^2|\lambda|^2}{8}-\frac{|c_1|^2}{2}\right)\frac{c_\phi}{16\pi M_{\widetilde{N}_+}}\nn\\
&&\times\bigg(2|A||M|\cos\left(\theta_A+\frac{\pi}{2}+\theta_M-\theta_b\right)\frac{|M|}{v_u^2}m_\text{eff}\nn\\
&&+|c_1||M|\lambda\sin\left(\theta_b-\theta_{c_1}+\theta_\lambda-\theta_M\right)\bigg)\nn\\
&&-\frac{4}{16\pi}\frac{M_{\widetilde{N}_-}^2-M_{\widetilde{N}_+}^2}{(M_{\widetilde{N}_-}^2-M_{\widetilde{N}_+}^2)^2+\Pi_{++}^2}\left(\frac{|c_1|^2}{2}-\frac{|\lambda|^2|M|^2}{8}\right)\frac{c_\phi}{16\pi M_{\widetilde{N}_-}}\nn\\
&&\bigg(2|M||A|\cos\left(\theta_M+\theta_A-\theta_b+\frac{\pi}{2}\right)\frac{|M|}{v_u^2}m_\text{eff}\nn\\
&&+|c_1||\lambda ||M|\sin\left(\theta_\lambda+\theta_b-\theta_{c_1}-\theta_M\right)\bigg)\Bigg]\nn\\
&&\times\Bigg[\frac{c_\phi}{8\pi M_{\widetilde{N}_+}}\left(\frac{|c_1|^2}{2}+\frac{|M|^2|\lambda|^2}{8}+\frac{1}{2}|c_1||M||\lambda|\cos\left(\theta_{c_1}-\theta_b+\theta_M-\theta_\lambda\right)\right)\nn\\
&&+\frac{c_\phi}{8\pi M_{\widetilde{N}_-}}\left(\frac{|M|^2|\lambda|^2}{8}+\frac{|c_1|^2}{2}-\frac{1}{2}|M||\lambda ||c_1|\cos\left(\theta_\lambda-\theta_M+\theta_b-\theta_{c_1}\right)\right)\Bigg]^{-1}
\eeqa
with the factors \cite{thermaleffects}
\beqa
c_F&=&\left(1-x_l-x_{\widetilde{H}}\right)\sqrt{\left(1+x_l-x_{\widetilde{H}}\right)^2-4x_l}\left(1-f_l^\text{eq}\right)\left(1-f_{\widetilde{H}} ^\text{eq}\right),\\
c_B&=&\sqrt{\left(1+x_H-x_{\widetilde{l}}\right)^2-4x_H}\left(1+f_H ^\text{eq}\right)\left(1+f_{\widetilde{l}} ^\text{eq}\right),\\
c_\phi&=&\sqrt{1-4x_\phi}\left(1+f_\phi ^\text{eq}\right)\left(1+f_\phi ^\text{eq}\right),\\
x_a&=&\frac{m_a(T)^2}{M^2},\ f^\text{eq}_a=\frac{1}{e^{E_a/T}\pm 1},
\eeqa
where the thermal masses are
\beqa
m_H(T)^2&=&T^2\left(\frac{3}{8}g_2^2+\frac{g_Y^2}{8}+\frac{3}{4}4\pi\alpha_t\right),\\
m_{\widetilde{H}}(T)^2&=&\frac{1}{2}T^2\left(\frac{3}{8}g_2^2+\frac{g_Y^2}{8}+\frac{3}{4}4\pi\alpha_t\right),\\
m_{\widetilde{l}}(T)^2&=&T^2\left(\frac{3}{8}g_2^2+\frac{g_Y^2}{8}\right),\\
m_l(T)^2&=&\frac{1}{2}T^2\left(\frac{3}{8}g_2^2+\frac{g_Y^2}{8}\right),\\
m_\phi(T)^2&=&m_\Phi^2+ m_{\Phi\text{soft}}^2+\lambda^2T^2.
\eeqa
The phases $\theta_M$ and $\theta_b$ can be rotated away and are set to zero and so $M$ is real and we can replace $|M|\rightarrow M$. The factors $c_{F,B,\phi}$ reflect the production thresholds of MSSM particles and DM boson $\phi$ imposed by energy conservation. Since we normalize the DM asymmetry to the DM boson production, the factors $c_\phi$ cancel and consequently there is no temperature dependence in $\eps_\text{DM}$. We have also kept the soft terms at the vertices for completeness, usually these are neglected \cite{ambgiurai}. 

\newpage

\section{Boltzmann equations}
\label{boltzmann}
\setcounter{equation}{0}
\setcounter{footnote}{0}

\begin{figure}[htb]
\centering
\includegraphics[scale=0.8]{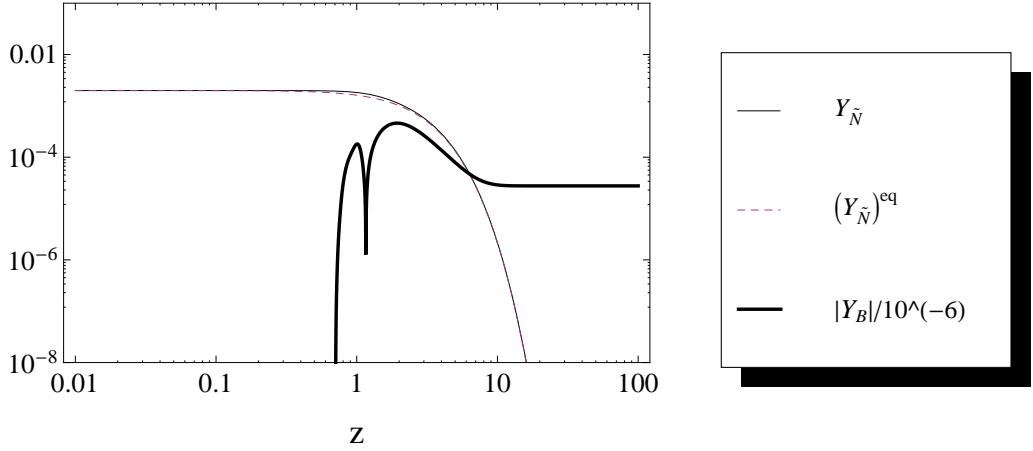}
\caption{Baryon $|Y_B|$ and sneutrino $Y_{\widetilde{N}}$ abundances with $M=10^5$ TeV, $m_\text{eff}=10^{-14}$ TeV, $A$=1 TeV, $\lambda=10^{-8}$ and $m_\Phi=0.1$ TeV.}
\label{babundance}
\end{figure}

\begin{figure}[htb]
\centering
\includegraphics[scale=0.8]{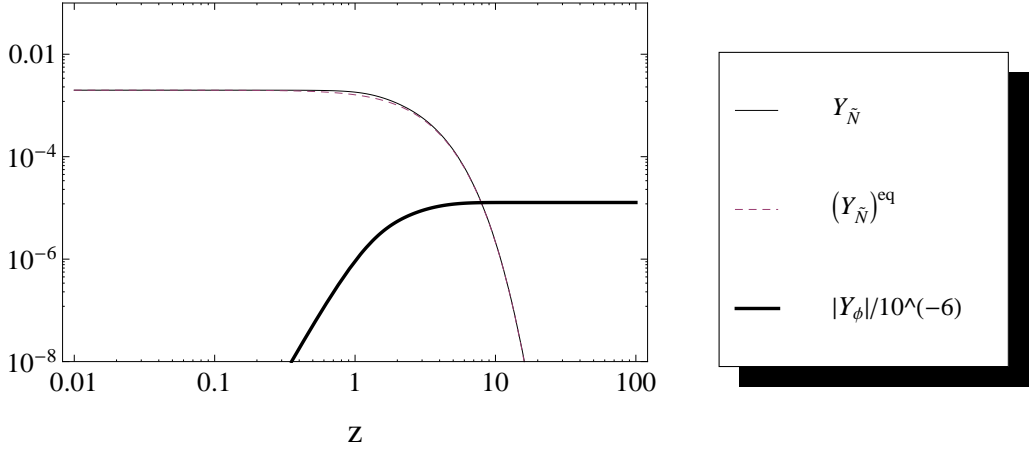}
\caption{Scalar DM $|Y_\phi|$ and sneutrino $Y_{\widetilde{N}}$ abundances with $M=10^5$ TeV, $m_\text{eff}=10^{-14}$ TeV, $A$=1 TeV, $|\lambda|=10^{-8}$, $m_\Phi=0.1$ TeV and $|c_1|=0.001$ TeV.}
\label{phiabundance}
\end{figure}

\begin{figure}[htb]
\centering
\includegraphics[scale=0.8]{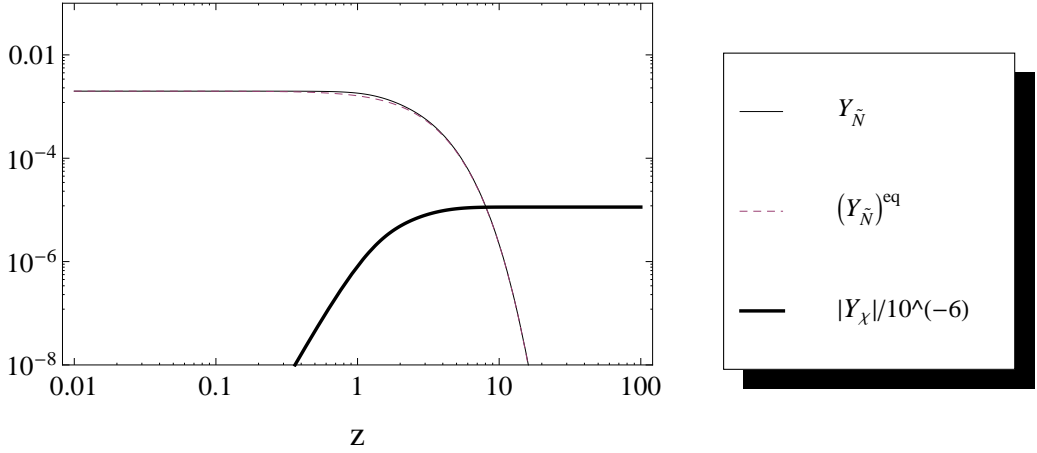}
\caption{Fermionic DM $|Y_\chi|$ and sneutrino $Y_{\widetilde{N}}$ abundances with $M=10^5$ TeV, $m_\text{eff}=10^{-14}$ TeV, $A$=1 TeV, $\lambda=10^{-8}$, $m_\Phi=0.1$ TeV and $|c_1|=0.001$ TeV.}
\label{chiabundance}
\end{figure}

We move on to solve the relic baryon and DM abundances this scenario generates. We assume the Yukawa couplings $m_\text{eff}$ and DM coupling $|\lambda|$ are sufficiently small and consider only processes with cross sections $\mc O(y^2)$ and $\mc O(\lambda^2)$. Due to this, DM annihilation processes are left out, which rules out ADM as the primary source of the DM energy density. Thus, we write the Boltzmann equations also for the total abundances of $\chi$ and $\phi$. The particle abundances are expressed in terms of the number densities normalized to entropy density $Y_i\equiv n_i/s$ and their evolution with changing $z\equiv M/T$ is to be determined. Our Boltzmann equations are %\cite{eungjinchun,bdp,slreview}
\beqa
\frac{dY_N}{dz}&=&-\left(Y_N-Y_N^\text{eq}\right)\left(D_N+D_{N\chi\phi}+4S_t^{(0)}+4S_t^{(1)}+4S_t^{(2)}+2S_t^{(3)}+4S_t^{(4)}\right),\\ 
\frac{dY_{\widetilde{N}}}{dz}&=& -\left(Y_{\widetilde{N}}-2Y_{\widetilde{N}}^\text{eq}\right)\Bigg[\frac{D_{\widetilde{N}}}{2} +D_{\widetilde{N}}^{(3)}+\frac{D_{\widetilde{N}\phi\phi}}{2} + \frac{D_{\widetilde{N}\chi\chi}}{2} + 3S_{22} \nn\\
&&+2\left(S_t^{(5)}+S_t^{(6)} + S_t^{(7)} + S_t^{(9)}\right) + S_t^{(8)}\Bigg],\\ 
\frac{dY_{\Delta L_\text{tot}}}{dz}&=& \eps_{L}(T)\left(Y_{\widetilde{N}}-2Y_{\widetilde{N}}^\text{eq}\right)\frac{D_{\widetilde{N}}}{2}\nn \\
&&-\frac{Y_{\Delta L_\text{tot}}}{Y_l^\text{eq}}\Bigg[\frac{D_N}{2}Y_N^\text{eq}+\frac{D_{\widetilde{N}}}{2}Y_{\widetilde{N}}^\text{eq}+Y_{\widetilde{N}}^\text{eq}D_{\widetilde{N}}^{(3)}+Y_{\widetilde{N}}\left(S_t^{(5)}+\frac{S_t^{(8)}}{2}\right)\nn\\
&&+Y_N\left(2S_t^{(0)}+S_t^{(3)}\right)+2Y_N^\text{eq}\left(S_t^{(1)}+S_t^{(2)}+S_t^{(4)}\right)\nn\\
&&+2Y_{\widetilde{N}}^\text{eq}\left(S_t^{(6)}+S_t^{(7)}+S_t^{(9)}\right)+S_{22}\left(2Y_{\widetilde{N}}^\text{eq}+\frac{Y_{\widetilde{N}}}{2}\right)\Bigg],\\ 
\frac{dY_{\Delta \phi}}{dz}&=&\eps_\text{DM}\left(Y_{\widetilde{N}}-2Y_{\widetilde{N}}^\text{eq}\right)\frac{D_{\widetilde{N}\phi\phi}}{2} -\frac{Y_{\Delta \phi}}{Y_\phi^\text{eq}}\left(\frac{D_{\widetilde{N}\phi\phi}+D_{\widetilde{N}\chi\chi}}{2}Y_{\widetilde{N}}^\text{eq}+\frac{D_{N \chi\phi}}{2}Y_N^\text{eq}\right),\\ 
\frac{dY_\chi}{dz}&=& -\left(\sqrt{1-\frac{4m_\chi^2}{M^2}}D_ {\widetilde{N}\chi\chi}+\sqrt{1+\frac{\left(m_\chi^2-m_\phi^2\right)^2}{M^4}-2\frac{m_\phi^2+m_\chi^2}{M^2}}D_{N\chi\phi}\right)\nn\\
&&\times\left(Y_{\widetilde{N}}-2Y_{\widetilde{N}}^\text{eq}\right),\\ 
\frac{dY_\phi}{dz}&=&-\left(\sqrt{1-\frac{4m_\phi^2}{M^2}}D_{\widetilde{N}\phi\phi}+\sqrt{1+\frac{\left(m_\chi^2-m_\phi^2\right)^2}{M^4}-2\frac{m_\phi^2+m_\chi^2}{M^2}}D_{N\chi\phi}\right)\nn\\
&&\times \left(Y_{\widetilde{N}}-2Y_{\widetilde{N}}^\text{eq}\right).
\eeqa
with $Y_{\widetilde{N}}^\text{eq}\approx Y_{\widetilde{N}_\pm}^\text{eq}$ and $Y_{\widetilde{N}}/2\approx Y_{\widetilde{N}_\pm}$ \cite{slreview}. Recall that we take into account $\mc O(y^2)$ and $\mc O(\lambda^2)$ decay rates and scattering diagrams involving the (s)top. The expressions for the decay $D$ and (s)top scattering rates $S$ can be found in appendix A. The solutions to these Boltzmann equations are shown in Figs. \ref{babundance}-\ref{chiabundance}. The baryon abundance is related to the (s)lepton abundance through
\beqa
Y_B=-\frac{8}{15}Y_{\Delta L_\text{tot}}.
\eeqa
We have chosen $m_\text{eff}=10^{-14}$ TeV, $M=10^5$ TeV, $M_h$=125 GeV, $|A|=1$ TeV, $|\lambda|=10^{-8}$, $|c_1|=0.001$ TeV and $m_\Phi=0.1$ TeV\footnote{If the heavy (s)neutrinos are supposed to be produced thermally, large $M$ implies reheating temperature $T_R>M$. In many inflatory supergravity models this high $T_R$ leads to overproduction of gravitinos. However, recently it has been proposed that with the present value of the Higgs mass about 125 GeV \cite{LHC} the reheating temperature constraint could be relaxed, $T_R\sim 10^{9}-10^{10}$ GeV, not in conflict with our scenario \cite{NTY}.}
The evolution of the $Y_B$ abundance is similar to those found in \cite{ambgiurai,slreview}. 

The DM abundances $Y_\phi$ and $Y_\chi$ are not sensitive to changes in the DM mass parameter $m_\Phi$ as long as it is a lot smaller than the heavy scale $M$ associated with the singlet (s)neutrino, $m_\Phi\ll M$. It is the coupling $\lambda$ and the positive mass dimension soft coupling $c_1$ that largely determine the order of magnitude in the DM abundances. With suitable choices of these couplings, the abundances $Y_\chi$ and $Y_\phi$ attain final values $\sim 10^{-11}$ suggesting that both the scalar and fermionic DM particles can be viable DM candidates. 

At small couplings $\lambda\sim10^{-8}$ and $c_1\sim 0.001$ TeV, the asymmetric component is extremely tiny with abundance $\sim 10^{-17}$ and does not play a role in DM energy density. However, an increase in both DM sector couplings causes the magnitude of $Y_{\Delta \phi}$ to rise and also compels to consider scatterings where DM particles are destroyed and MSSM particles are produced. So, in order to have ADM as the primary source of all DM, it is crucial to find a balance between sufficient DM production through asymmetric decays and scattering processes 
\beqan
&&\phi\phi\rightarrow \tilde L_\alpha H_u,\ \phi\phi\rightarrow L_\alpha \widetilde{H}^c_u, \\
&&\chi\chi^c\rightarrow \tilde L_\alpha H_u,\ \chi\chi^c\rightarrow L_\alpha \widetilde{H}^c_u,\\
&&\chi\phi\rightarrow L_\alpha H_u, \chi\phi\rightarrow \tilde L_\alpha \widetilde{H}_u^c, 
\eeqan
where $\chi^c$ denotes the right-handed fermion obtained from $\chi$ through $\bar\chi^{\dot{\alpha}}=\epsilon ^{\dot{\alpha}\dot{\beta}}\left(\chi_\beta\right)^*$, where $\epsilon ^{\dot{\alpha}\dot{\beta}}=\left(-i\sigma^2\right)^{\dot{\alpha}\dot{\beta}}$. We believe it is possible to create conditions where the DM particles annihilate and the asymmetric component remains and forms a part of final DM abundance although this scenario entails a different parameter space in terms of DM couplings and other parameter inputs. We are studying this alternative in an upcoming paper. 

\begin{figure}[htb]
\centering
\includegraphics[scale=0.7]{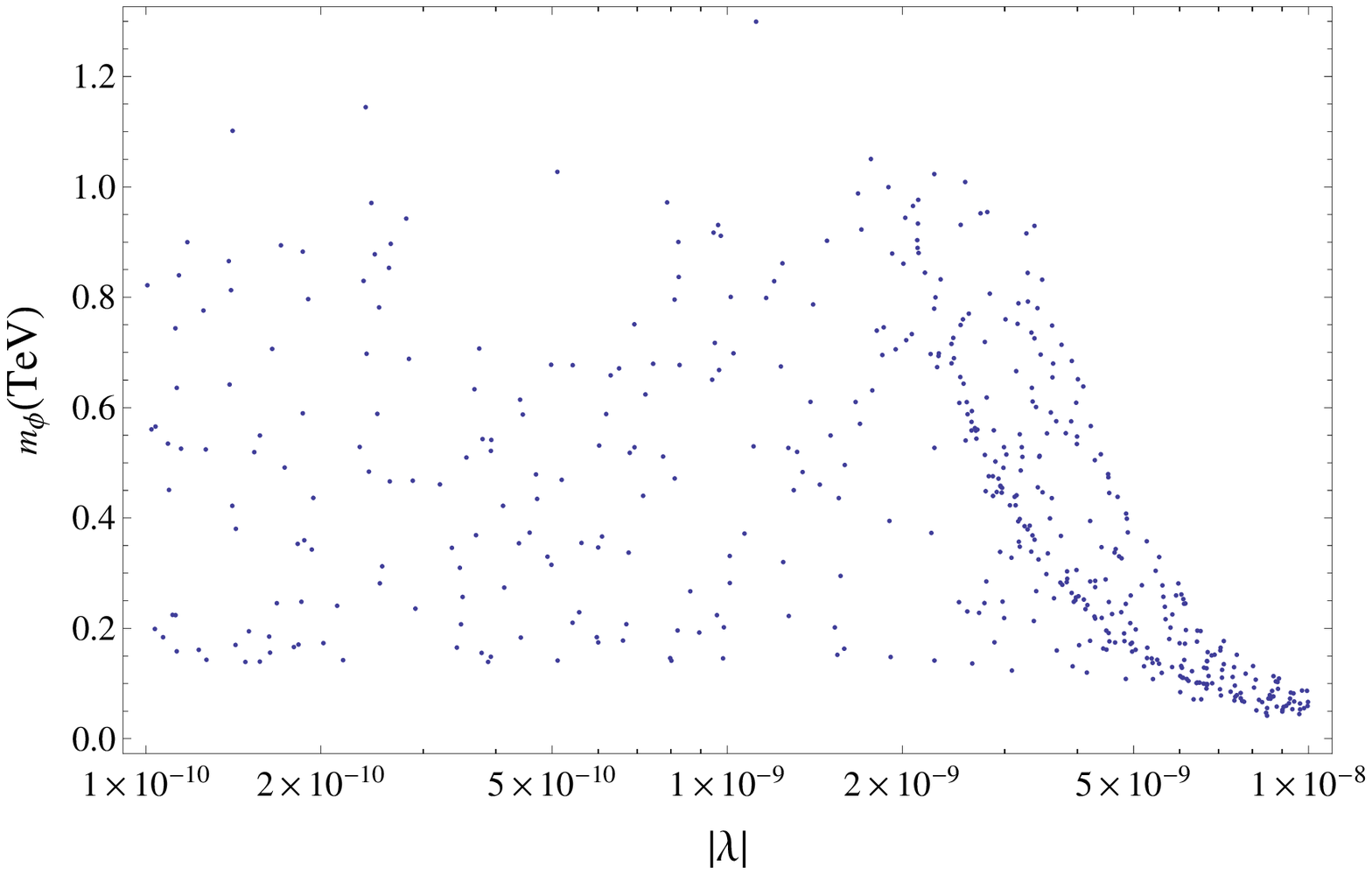}
\caption{Scatter plot of $m_\phi$ vs $|\lambda|$ with the present baryon and DM abundances as constraints, Eqs. (\ref{dmmasscoupling}) and (\ref{baryonconstraint}).}
\label{mphivsl}
\end{figure}

\begin{figure}[htb]
\centering
\includegraphics[scale=0.7]{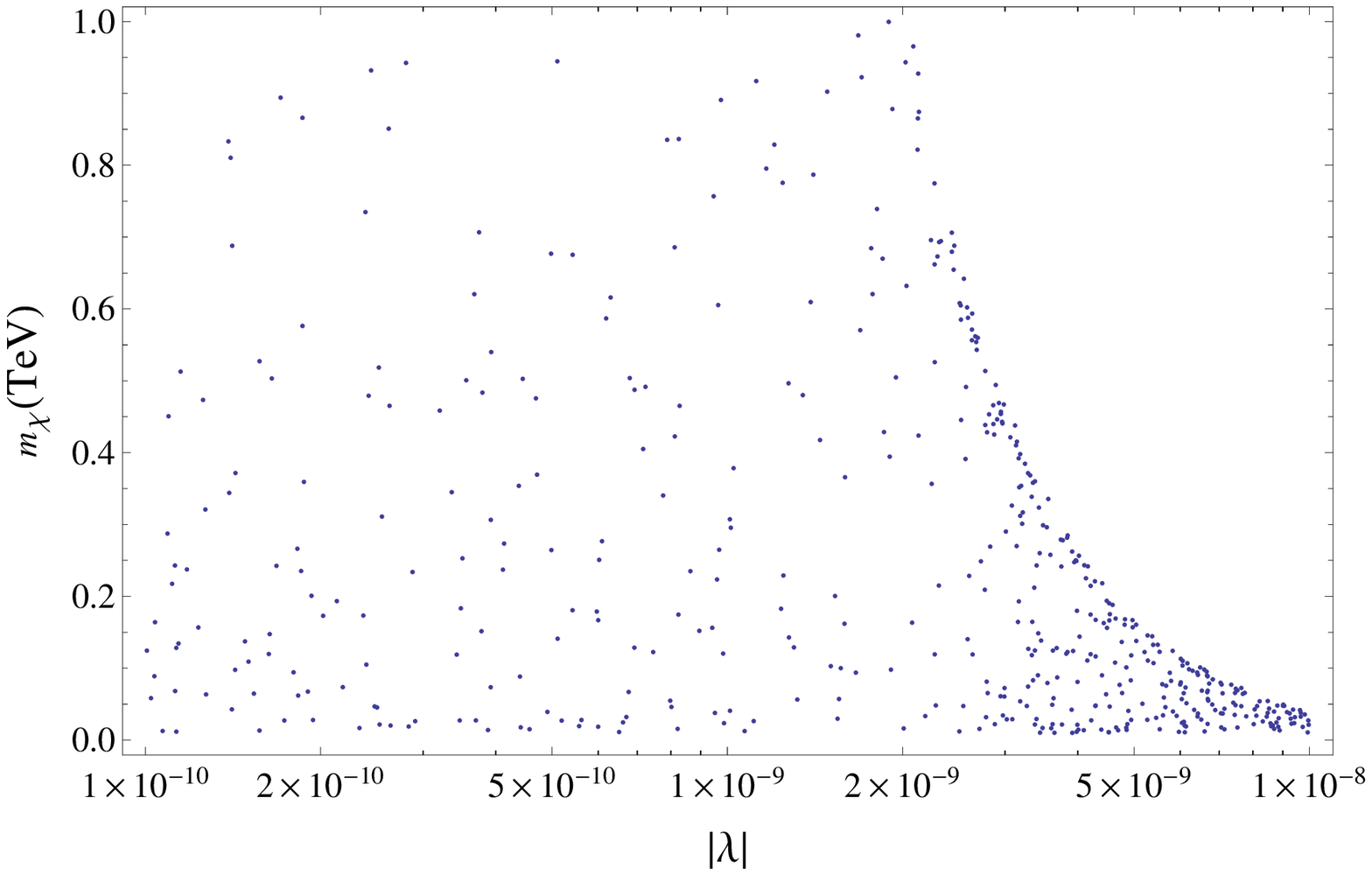}
\caption{Scatter plot of $m_\chi$ vs $|\lambda|$ with the present baryon and DM abundances as constraints, Eqs. (\ref{dmmasscoupling}) and (\ref{baryonconstraint}).}
\label{mchivsl}
\end{figure}

We have explored the allowed parameter space and made scatter plots of the DM masses versus the DM coupling $|\lambda|$, Figs. \ref{mphivsl}-\ref{mchivsl}. The fixed parameters are $m_\text{eff}=10^{-14}$ TeV, $M=10^5$ TeV and $A=1$ TeV while $|\lambda|$, $|c_1|$, $m_\Phi$ and $m_{\Phi\text{soft}}$ are varied as shown in Table 1. The range for the soft parameter $|c_1|$ has been chosen so that $|c_1|\sim |\lambda|M$. The DM coupling $10^{-10}<|\lambda|<10^{-8}$ is the most favorable region of producing the correct DM abundance which is why this particular region is studied. The DM mass parameters are chosen to be in the weak scale and even in the 10 GeV range which brings the typical ADM regime closer. 
\begin{table}
\caption{The DM sector parameters and their scanned values.}
\label{parameterscan}
\begin{center}
\begin{tabular}{|c|c|} 
\hline
Parameters & Scanned values \\
\hline
$|\lambda|$ & $10^{-10}-10^{-8}$ \\
\hline
$|c_1|$ & $10^{-5}-10^{-3}$ TeV\\
\hline
$m_\Phi$ & 0.01-1 TeV \\
\hline
$m_{\Phi\text{soft}}^2$ & $10^{-4}$-1 TeV$^2$ \\
\hline
\end{tabular}
\end{center}
\end{table}
The required DM abundances come from the relation
\beqa \label{dmmasscoupling}
m_\chi |Y_{\chi}|+m_\phi |Y_\phi|\lesssim\frac{3}{4}\frac{\Omega_\text{DM}}{\Omega_\gamma}T
\eeqa
and the baryon abundance has to satisfy 
\beqa \label{baryonconstraint}
|Y_B|\lesssim 8.8\times 10^{-11}
\eeqa
which we have used as constraints in extracting the allowed parameter space. Known parameter values from observations are $T=2.725$ K, $\Omega_\text{DM}=0.11/h^2$ and $\Omega_\gamma=2.47\times 10^{-5}/h^2$. We have taken the lower limit in (\ref{baryonconstraint}) to be 30 $\%$ of the observed baryon abundance while in (\ref{dmmasscoupling}) the lower limit is 90 $\%$ of the observed value. Figs. \ref{mphivsl}-\ref{mchivsl} reflect the hyperbola-like boundary set by Eq. (\ref{dmmasscoupling}), and there is an indication of slight preference of $|\lambda|$ values in the $\sim 10^{-9}$ ballpark. A similar observation of the favorable magnitude of $|\lambda|$ can be made from a plot with the left-hand side of (\ref{dmmasscoupling}) plotted against $|\lambda|$.

\begin{figure}[htb]
\centering
\includegraphics[scale=0.7]{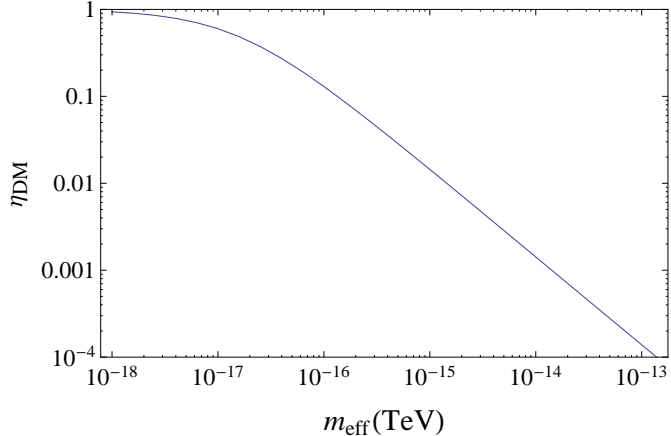}
\caption{The efficiency factor $\eta_\text{DM}$ for asymmetric DM production vs $m_\text{eff}$.}
\label{etadmvsmeff}
\end{figure}

\begin{figure}[htb]
\centering
\includegraphics[scale=1]{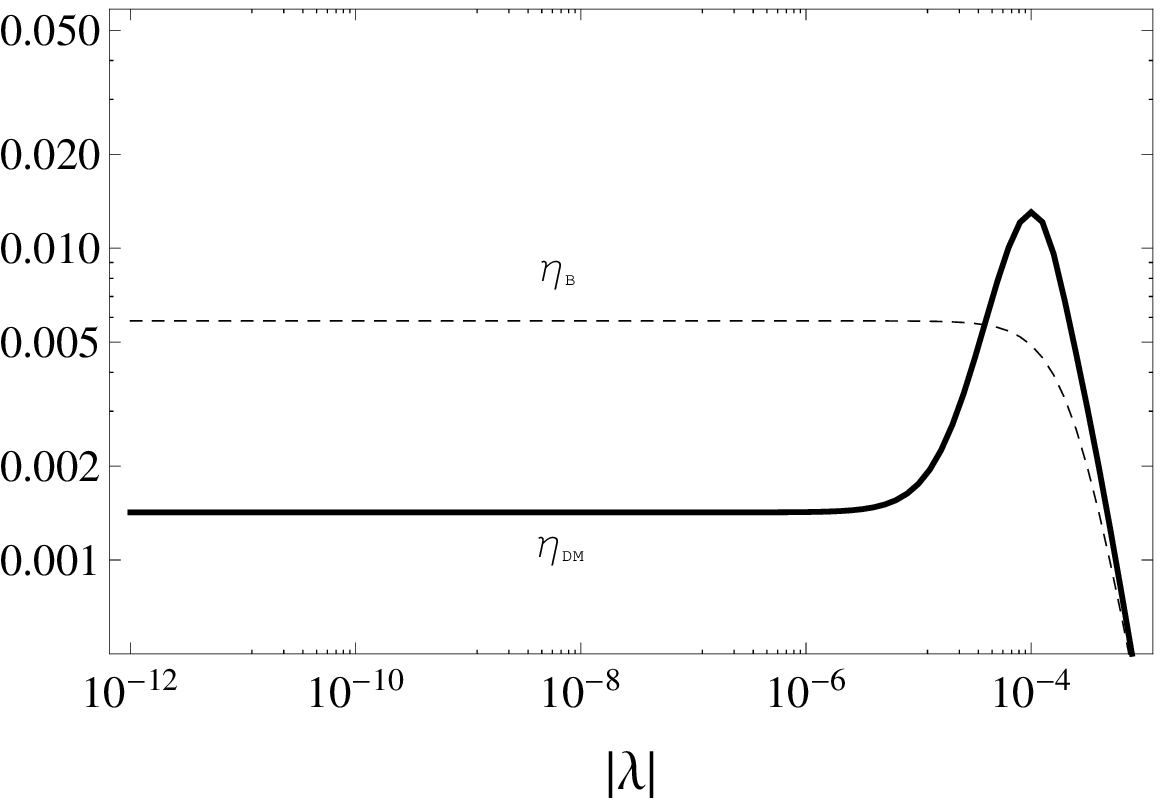}
\caption{The efficiency factors $\eta_B$ and $\eta_\text{DM}$ vs $|\lambda|$ for $m_\text{eff}=10^{-14}$ TeV.}
\label{etasvslambda}
\end{figure}

We have also determined the dependence of efficiency factors related to baryon and ADM production on the Yukawa and DM couplings. The efficiency factors $\eta_B$ and $\eta_\text{DM}$ are shown in Figs. \ref{etadmvsmeff}-\ref{etasvslambda} and they are given by
\beqa
\eta_{B,\text{DM}}=\left|\frac{Y_{\Delta L_\text{tot},\Delta \phi}}{2\eps_{L,\text{DM}}Y_{\widetilde{N}}^\text{eq}(z\approx 0)}\right|.
\eeqa
The ADM production efficiency falls with rising $m_\text{eff}$ while with increasing $|\lambda|$ $\eta_\text{DM}$ increases and after reaching a maximum a steep decline ensues. The fall in $\eta_\text{DM}$ is due to the fact that while $|\lambda|$ increases, DM production climbs up but sooner or later washout processes overcome the production. This observation is also made in \cite{eungjinchun}. Of course a more accurate prediction of the DM efficiency factor can be made when the scenario better fulfills the criteria for the existence of ADM, which would mean larger DM couplings $|\lambda|$ (and $|c_1|$) and the inclusion of DM annihilations into MSSM particles.


\section{Conclusions}
\label{paatelmat}

We have studied soft leptogenesis with an additional singlet DM superfield that couples to the singlet neutrino superfield. This framework allows the production of DM through the heavy (s)neutrino decays. In the weak coupling regime we have neglected suppressed higher order scattering processes, which does not make the ADM scenario viable due to the lack of DM annihilation channels. The absence of annihilations makes the DM produced in the tree level decays $\widetilde{N}_\pm\rightarrow \chi\chi^c$, $\widetilde{N}_\pm\rightarrow \phi\phi$ and $N\rightarrow \chi\phi$ stable.
%provides an alternative in case the LSP paradigm for DM fails for some reason

It turns out that sufficient levels of baryons and DM can be produced in accordance with observations without the need for ADM. This is supported by the evolution of the final abundances $Y_B$ and $Y_{\chi,\phi}$ and parameter space related to the DM sector. We have produced scatter plots depicting the overall dependence of DM masses $m_{\chi,\phi}$ versus the DM coupling $\lambda$ with cosmological observations of the baryonic matter and DM energy densities as constraints, Eqs. (\ref{dmmasscoupling}) and (\ref{baryonconstraint}). The DM coupling $\lambda$ and its soft counterpart $c_1$ are independent from the MSSM and soft leptogenesis parameters and thus their magnitudes are fixed by the present DM abundance, and also the baryon abundance through loop effects in $\eps_L$ in Eq. (\ref{epsl}). The plots in Figs. \ref{mphivsl} and \ref{mchivsl} exhibit the fact that the observed DM energy density restricts the allowed region by making a hyperbola-like boundary.

For future studies, it would be interesting to see what is the correct parameter space that allows for ADM in this scenario. We believe this happens once we take into account the annihilation processes for the DM particles. The leading annihilation channel for scalar DM $\phi\phi\rightarrow \tilde L_\alpha H_u$ comes from the $F$-term potential. Other channels are mediated by $N$ and $\widetilde{N}_\pm$. Thus, a stronger coupling of the DM sector can not only lead to higher levels of the asymmetric abundance $Y_{\Delta \phi}$ but also makes the DM annihilations significant which is a necessary precursor for the existence of ADM. 

{\bf Acknowledgments} The work of H.K. was supported by Alfred Kordelin Foundation.

\appendix 

\section{Decay and scattering rates}
\setcounter{equation}{0}
\setcounter{footnote}{0}

The (s)top scattering rates are given by \cite{bdppedestrians,slreview,plumacher}
\beqa
S_t^{(0)}&=&\frac{K_S}{6}P_0(z),\ S_t^{(1)}=\frac{K_S}{6}P_1(z)\nn,\\
S_t^{(2)}&=&\frac{K_S}{6}P_2(z),\ S_t^{(3)}=\frac{K_S}{6}P_3(z)\nn,\\
S_t^{(4)}&=&\frac{K_S}{6}P_4(z),\ S_t^{(5)}=\frac{K_S}{6}P_5(z)\nn,\\
S_t^{(6)}&=&\frac{K_S}{6}P_6(z),\ S_t^{(7)}=\frac{K_S}{6}P_7(z)\nn,\\
S_t^{(8)}&=&\frac{K_{SS}}{6}P_8(z),\ S_t^{(9)}=\frac{K_{SS}}{6}P_9(z)\nn,\\
S_{22}&=&3\alpha_t\frac{M}{8\pi^2H(z=1)}\frac{M}{v_u^2}\frac{m_\text{eff}}{z}\frac{K_1(z)}{K_2(z)},
\eeqa
where 
\beqa
K_S&=&18\alpha_t\frac{M}{v_u^2}m_\text{eff}\frac{v_u^2}{M}\frac{3\sqrt{5}M_\text{Pl}}{16^2\pi^2\pi^{5/2}\sqrt{g_*}v_u^2},\\
K_{SS}&=&18\alpha_t\frac{M}{v_u^2}m_\text{eff}\frac{M^2+|A|^2}{M^2}\frac{v_u^2}{M}\frac{3\sqrt{5}M_\text{Pl}}{16^2\pi^2\pi^{5/2}\sqrt{g_*}v_u^2}
\eeqa
The reaction rates are found by integrating the functions ($x\equiv \psi/z^2$)
\beqa
f_3(x)&=&\frac{(x - 1)^2}{x^2}\nn,\\
f_4(x)&=&\frac{x -1}{x}\left [\frac{x-2+2a_h}{x-1+a_h}+\frac{1 - 2 a_h}{x - 1}\log\left (\frac{x - 1 + a_h}{a_h}\right )\right ]\nn,\\
f_0(x)&=&\frac{1}{2}\frac{x^2 - 1}{x^2}\nn,\\
f_1(x)&=&\frac{x - 1}{x}\left[-\frac{2x - 1 + 2a_h}{x - 1 + a_h}+\frac{x + 2a_h}{x - 1}\log\left(\frac{x - 1 + a_h}{a_h}\right)\right ]\nn,\\
f_2(x)&=&\frac{x - 1}{x}\left[-\frac{x - 1}{x - 1 + 2a_h}+\log\left(\frac{x - 1 + a_h}{a_h}\right)\right]\nn,\\
f_5(x)&=&\frac{1}{2}\frac{(x - 1)^2}{x^2 }\nn,\\
f_6(x)&=&\frac{x - 1}{x}\left[-2 + \frac{x - 1 + 2a_h}{x - 1}\log\left(\frac{x - 1 + a_h}{a_h}\right)\right]\nn,\\
f_7(x)&=&-\frac{x - 1}{x - 1 + 2a_h} + \log\left(\frac{x - 1 + a_h}{a_h}\right)\nn,\\
f_8(x)&=&\frac{x - 1}{x^2}\nn,\\
f_9(x)&=&\frac{1}{x}\left[-\frac{x - 1}{x - 1 + a_h}+\log\left(\frac{x - 1 + a_h}{a_h}\right)\right],
\eeqa
where $a_h=(M_h/M)^2$ with $M_h$ standing for the mass of the Higgs boson. The integrals determining the reaction rates are of the form
\beqa
P_i(z)&=&\frac{K_2(z)}{z^2}\int_{z^2}^\infty d\psi f_i(\psi)\sqrt{\psi}K_1\left(\sqrt{\psi}\right).
%P_3(z)&=&\frac{K_2(z)}{z^2}\int_{z^2}^\infty d\psi f_3(\psi)\sqrt{\psi}K_1\left(\sqrt{\psi}\right),\\
%P_4(z)&=&\frac{K_2(z)}{z^2}\int_{z^2}^\infty d\psi f_4(\psi)\sqrt{\psi}K_1\left(\sqrt{\psi}\right),\\
%P_0(z)&=&\frac{K_2(z)}{z^2}\int_{z^2}^\infty d\psi f_0(\psi)\sqrt{\psi}K_1\left(\sqrt{\psi}\right),\\ 
%P_1(z)&=&\frac{K_2(z)}{z^2}\int_{z^2}^\infty d\psi f_1(\psi)\sqrt{\psi}K_1\left(\sqrt{\psi}\right),\\
%P_2(z)&=&\frac{K_2(z)}{z^2}\int_{z^2}^\infty d\psi f_2(\psi)\sqrt{\psi}K_1\left(\sqrt{\psi}\right),\\
%P_5(z)&=&\frac{K_2(z)}{z^2}\int_{z^2}^\infty d\psi f_5(\psi)\sqrt{\psi}K_1\left(\sqrt{\psi}\right),\\
%P_6(z)&=&\frac{K_2(z)}{z^2}\int_{z^2}^\infty d\psi f_6(\psi)\sqrt{\psi}K_1\left(\sqrt{\psi}\right),\\
%P_7(z)&=&\frac{K_2(z)}{z^2}\int_{z^2}^\infty d\psi f_7(\psi)\sqrt{\psi}K_1\left(\sqrt{\psi}\right),\\
%P_8(z)&=&\frac{K_2(z)}{z^2}\int_{z^2}^\infty d\psi f_8(\psi)\sqrt{\psi}K_1\left(\sqrt{\psi}\right),\\
%P_9(z)&=&\frac{K_2(z)}{z^2}\int_{z^2}^\infty d\psi f_9(\psi)\sqrt{\psi}K_1\left(\sqrt{\psi}\right).
\eeqa
The (s)neutrino decay rates are given by
\beqa
D_N&=&\frac{M}{v_u^2}m_\text{eff}\frac{M}{(4\pi)}z\frac{K_1(z)}{K_2(z)}\nn,\\
D_{\widetilde{N}}&=&\left(\frac{1}{8\pi M}\left(M^2+|A|^2\right)+\frac{1}{8\pi}M\right)\frac{M}{v_u^2}m_\text{eff}z\frac{K_1(z)}{K_2(z)}\frac{1}{H(z=1)}\nn,\\
D_{\widetilde{N}\phi\phi}&=&\frac{1}{4\pi M}\left(\frac{|c_1|^2}{2}+\frac{M^2|\lambda|^2}{8}\right)z\frac{K_1(z)}{K_2(z)}\frac{1}{H(z=1)}\nn,\\
D_{\widetilde{N}\chi\chi}&=& \frac{1}{8\pi}M|\lambda|^2z\frac{K_1(z)}{K_2(z)}\frac{1}{H(z=1)}\nn,\\
D_{\widetilde{N}\chi\phi}&=&\frac{1}{8\pi}M|\lambda|^2z\frac{K_1(z)}{K_2(z)}\frac{1}{H(z=1)}\nn,\\
D_{\widetilde{N}}^{(3)}&=&\frac{3\alpha_t}{64\pi^2}M\frac{M}{v_u^2}\frac{m_\text{eff}}{H(z=1)}z\frac{K_1(z)}{K_2(z)}.
\eeqa

\end{document}